\documentclass[prl,twocolumn, preprintnumbers,showkeys, amssymb]{revtex4}
\usepackage{graphicx}
\usepackage{amssymb}
\usepackage{amsmath}
\usepackage{bm}
\begin{document}
\setcounter{page}{0}
\title{Comment on ``Critical and slow dynamics in a bulk metallic glass exhibiting strong random magnetic anisotropy" [Appl. Phys. Lett. 92, 011923 (2008)] }
\author{Ha M. Nguyen }

\email{nmha@ess.nthu.edu.tw}
\author{Pai-Yi Hsiao}
\email{pyhsiao@ess.nthu.edu.tw}
\affiliation{Department of Engineering and System Science, National Tsing Hua University, Hsinchu, Taiwan 30013, R.O.C}
\date{\today}
\begin{abstract}
In this comment, by using Monte Carlo simulation, we show that the perpendicular shift of hysteresis loops reported in the commented work is nothing special but simply due to the fact that the range of field does not surpass the reversible field beyond which the two branches of the loop merge. If the reversible field is exceeded, the shift is no longer observed. Moreover, we point out that even using a small range of field, the shift will not be observed if the observation time is long enough for the reversible field to drop within the range.
\end{abstract}
\maketitle
In a recent work, Luo {\it et al.} \cite{ref1} have presented the perpendicular shift of hysteresis loops (in Fig. 1(d) of Ref. \cite{ref1}) of a Dy-based $\rm Dy_{40}\rm Al_{24}\rm Co_{20}\rm Y_{11}\rm Zr_{5}$ alloy, a bulk metallic glass with strong random magnetic anisotropy (RMA), after field-of-500-Oe cooling the glass to 2 K, the temperature well below the spin-glass (SG) transition point $T_{\rm g} = 16.6$ K. We show that it was irrelevant to use this shifting behavior as a peculiarity of the alloy to contrast against the exchange bias intrinsic to the domain states in Cu-Mn and Ag-Mn GSs \cite{ref2}. This is because the shift is nothing special but simply due to an experimental fact that the range of measuring fields, $\pm H_{\rm m}$, does not exceed the reversible fields, $\pm H_{\rm rev}$, beyond which the two branches of the loop merge. When measuring the hysteresis loop with $H_{\rm m}\geq H_{\rm rev}$ the shift is no longer observed.

\begin{figure}
       \begin{center}
         \resizebox{80 mm}{!}{\includegraphics{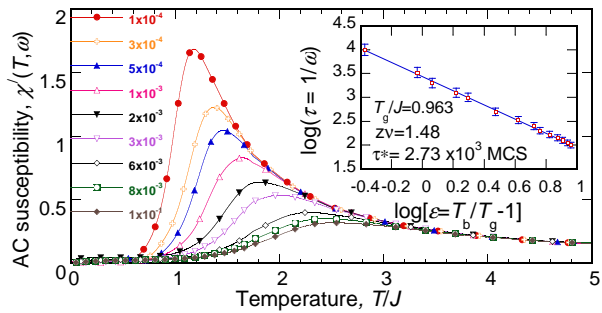}}         
        \end{center}
          \caption{Temperature dependence of $\chi^{\prime}(T,\omega)$ at various frequencies. The inset shows the law of critical slowing-down dynamics of a magnetic glass.}
       \label{fig1}  
\end{figure}

To support our argument, in this comment, we conduct a Monte Carlo (MC) simulation upon the so-called RMA model \cite{ref3}. The Hamiltonian of the model has been described in detail in our recent reports \cite{ref4,ref5}. Here, chosen is only one case of the anisotropy-to-exchange ratio $D/J=10$ which has been shown to be strong RMA of a speromagnet \cite{ref5,ref6}. For the sake of qualitative illustration, a moderate computation is carried out for simple-cubic-lattice systems of $10\times10\times10$ Heisenberg spins in which each point of data is averaged over 50 independent realizations. As calculated by using the same technique as that shown in Ref. \cite{ref5}, the real part, $\chi^{\prime}(T,\omega)$, of the ac susceptibility in Fig. \ref{fig1} exhibits the critical slowing-down dynamics, $\tau =\tau^{*}(T_{\rm b}/T_{\rm g}-1)^{-z\nu}$, at $T_{\rm g}/J\simeq 0.963$ with $z\nu \simeq 1.48$. The scaled plot for this law can be seen in the inset of Fig. \ref{fig1}. $T_{\rm b}$'s in the equation are the temperatures at the peaks of $\chi^{\prime}(T,\omega)$ corresponding to various frequencies in the range $1\times10^{-4}$ MCS$^{-1}$ $\le \omega \le 1\times10^{-2}$ MCS$^{-1}$ and $\tau =1/\omega$, where MCS stands for Monte Carlo step (MCS) per spin. Therefore, this dynamical behavior is qualitatively similar to that presented in Fig. 2(d) of Ref. \cite{ref1}.

\begin{figure}
       \begin{center}
         \resizebox{80 mm}{!}{\includegraphics{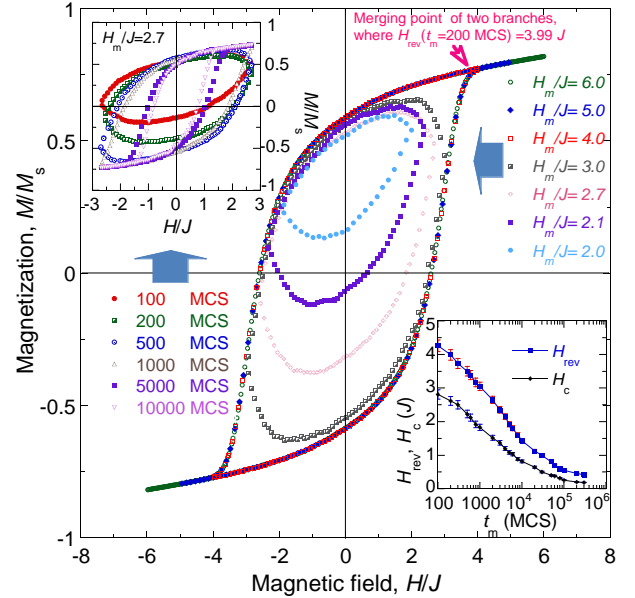}}         
        \end{center}
          \caption{Hysteresis loops measured at $T/J=0.75$ with cooling field $H/J=1.0$. The upper inset shows loops for $H_{\rm m}/J=2.7$ with various windows of observation time. The lower inset presents the dependences of the reversible field and the coercivity on observation time.}
       \label{fig2}  
   \end{figure}
To calculate the hysteresis loop, we use Metropolis technique in our MC simulation, the detail of the calculation has been shown in Refs. \cite{ref7,ref8}. Fig. \ref{fig2} presents the loops measured at temperature $T/J=0.75$ after field-cooling with magnetic field $H/J=1.0$. In this case, the observation time is $t_{\rm m}=200$ MCS which corresponds to the reversible field $H_{\rm rev}/J\simeq 3.99$ (the field at the merging point indicated by the pink-colored arrow and text in Fig. \ref{fig2}). Apparently, the centers of those loops that are measured in the ranges of measuring field  not surpassing $H_{\rm rev}$ (e.g., those of $H_{\rm m}/J=2.0, 2.1, 2.7$, and $3.0$) shift upward with decreasing $H_{\rm m}$. Qualitatively, these are what Luo {\it et al.} have observed in their measurements \cite{ref1}. On the other hand, all of the loops with $H_{\rm m} > H_{\rm rev}$ (i.e., $H_{\rm m}/J=4.0, 5.0$, and $6.0$) coincide thoroughly and are symmetric about the origin, that is, the shift is not observed any more. Therefore, in their experiment if Luo {\it et al.} had been able to determine $H_{\rm rev}$ priorly so that they could have extended ranges of measuring field beyond $H_{\rm rev}$, and then the shift would have not been observed either.  Moreover, we would like to emphasize that the window of observation time is an important key in the magnetization measurements of magnetic glasses. Actually, $H_{\rm rev}$ decreases with observation time as shown in the lower inset of Fig. \ref{fig2}. This reduction accounts for what is observed in the upper inset of Fig. \ref{fig2}. If the range of measuring field is fixed, e.g., $H_{\rm m}/J=2.7$, the upward shift is observed with the window of observation time getting narrower and narrower. In contrast, for $t_{\rm m}\geq 5000$ MCS, because $H_{\rm rev}$ drops within the range of measuring field, i.e., $H_{\rm rev}\leq H_{\rm m}=2.7 J$, symmetric loops are observed as usual, except for the reduction of the loop's area as a result of the decrease of the coercivity, $H_{\rm c}$, with observation time [see the lower inset of Fig. \ref{fig2}]. Again, we anticipate that even in a narrow range of field, Luo {\it et al.} still observe usual symmetric loops if the observation time in their experiment is long enough so that $H_{\rm rev}$ stays within the range. Finally, it is worthnoting that although in reality $H_{\rm rev}$ and $H_{\rm c}$ may decrease with observation time much slower for one magnetic glass than for another they would vanish with infinitely long observation time. In this approach, all hysteresis loops would be expected to reduce to reversible anhysteretic ones measured above the blocking temperature of ensembles of magnetic nanoparticles or nanoclusters \cite{ref7, ref8,ref9,ref10}. Unfortunately, nobody can be patient enough to wait for such long experiments!

In summary, it is pointed out from our MC simulation that when measuring the hysteresis loop of magnetic glasses, especially hard magnetic ones \cite{ref7}, care must be taken in choosing the range of measuring field and the window of observation time, otherwise confusion is very likely reported as an output of the experiment. We hope that our comment will stimulate the authors of Ref. \cite{ref1} to reexamine their measurements in order to have a final proper conclusion.

This work was financially supported by the National Science Council of Taiwan, R.O.C, under  Grant No. NSC 97-2112-M-007-007-MY3. The computing resources of the National Center for High-Performance Computing under the project ``Taiwan Knowledge Innovation National Grid" are acknowledged.

Ha M. Nguyen would like to dedicate this work as a special gift to Professors N. X. Phuc and L. V. Hong, his former advisors, and Yen L. T. Nguyen, his beloved mother, on their 60th anniversaries of birthday in 2009.

\end{document}